\documentclass[aip,jcp,amsmath,amssymb,preprint,floatfix]{revtex4-1}
\usepackage{graphicx}
\usepackage{dcolumn}
\usepackage{bm}
\usepackage[usenames,dvipsnames]{xcolor}

\begin{document}

\title {Note: Formation of the nematic splay-bend in two-dimensional systems of bow-shaped particles}

\author{Pawe\l{} Karbowniczek}
\affiliation{Institute of Physics, Cracow University of Technology, ul.\ Podchor\c{a}\.{z}ych 1, 30-084, Krak\'{o}w, Poland.}

\date{\today}

\maketitle

Recently, Tavarone \textit{et al.} \cite{tavarone} discussed phase behavior of zig-zag and bow-shaped particles composed of three needles. The authors presented very interesting results of extensive Monte Carlo simulations with periodic boundary conditions in the constant-NVT and the constant-NPT ensembles. In addition to isotropic, nematic, and smectic phases,  they identified a modulated nematic, which is actually the nematic splay-bend phase ($N_{SB}$), long-anticipated for bent-core systems \cite{dozov}. They also described isotropic-nematic and nematic-smectic transitions using Density Functional Theory in mean-field approximation. The authors, however, did not provided a theoretical description of the $N_{SB}$. Here, we present a simple theory of a phase transition to the $N_{SB}$ phase to fill the gap.

\begin{figure}[!htb]
\begin{center}
\includegraphics[width=0.45\columnwidth]{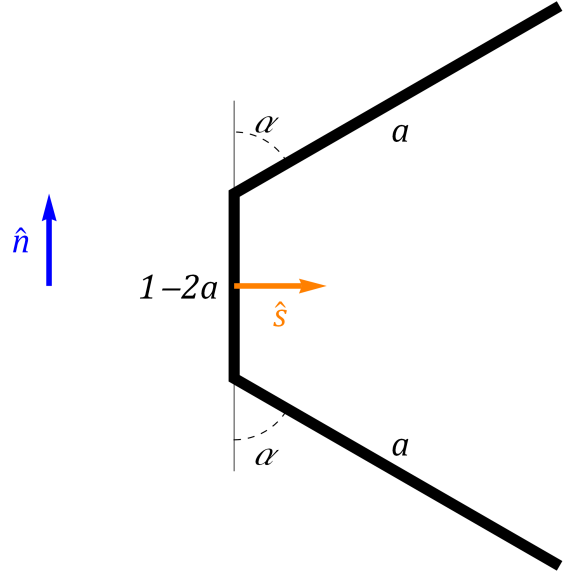}
\end{center}
\caption{Parametrization of studied bow-shaped particles. Central line segment of length $1-2a$ is attached at angle $\alpha$ to two side segments, each of length equal to $a$. Steric dipole $\mathbf{\hat{s}}$ of the particle is perpendicular to the local director $\mathbf{\hat{n}}$.}
\label{figm1}
\end{figure}

\begin{figure}[!htb]
\begin{center}
\includegraphics[width=0.9\columnwidth]{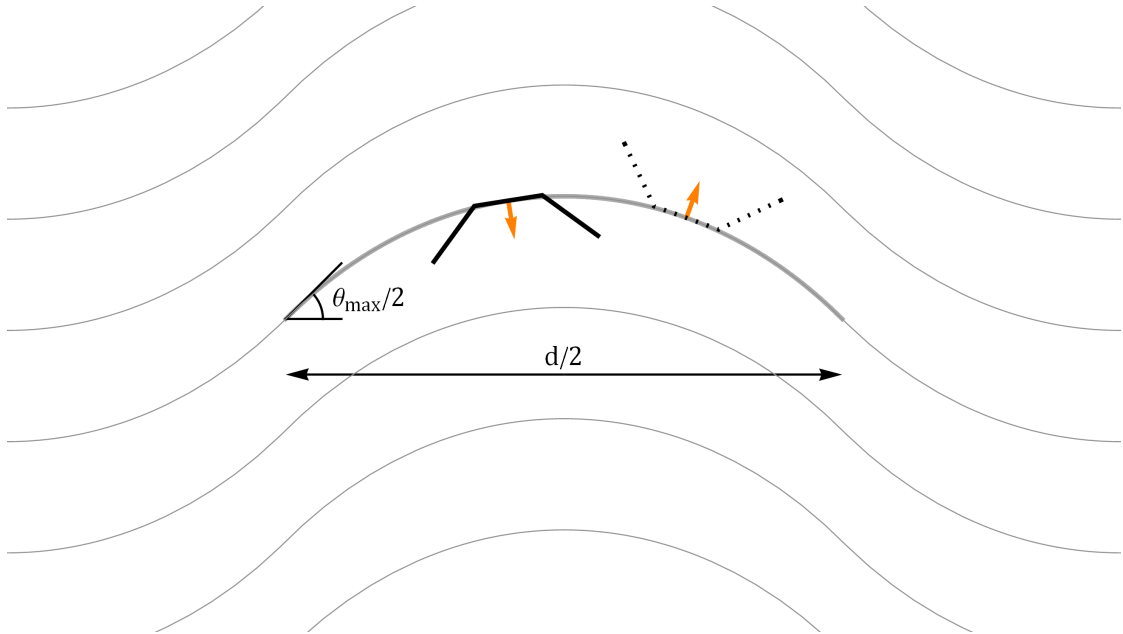}
\end{center}
\caption{Parametrization of the $N_{SB}$ phase. Local director is always tangent to the curves shown.}
\label{fig0}
\end{figure}

We use Onsager-type Density Functional Theory with perfect order approximation, in which orientational degrees of freedom are reduced to two possible perpendicular orientations of the steric dipole with respect to the local director. This assumption is valid for highly ordered phases, like the nematic splay-bend.
Similarly to our previous analysis \cite{karbowniczek}, for a proper description of the $N_{SB}$ structure we use Meyer parametrization \cite{meyer}. In the $N_{SB}$ we assume that the bow-shaped particles (see Fig. \ref{figm1}), for which this phase occurs, can rotate between $-\theta_{max}/2 $ and $\theta_{max}/2 $ for $0\le x <d/2$ (see Fig. \ref{fig0}), where $d$ is the period of the structure.
Using perfect local order approximation probability distribution function of the $N_{SB}$ is given by
\begin{equation}
  P(s,{\mathbf{\hat n}}(x)) = \frac{1}{2 S} +
  \frac{1}{2S} \langle s \rangle s \delta[{\mathbf{\hat{s}}}(x) +\{ \sin(\phi(x)), \cos(\phi(x))\} ],
\end{equation}
where $S$ is the surface area, $s$ is the orientational degree of freedom representing steric dipole ($-1$ for particles pointing to the left and $+1$ for particles pointing to the right), ${\mathbf{\hat{s}}}(x)$ is the local polarization perpendicular to the local director ${\mathbf{\hat{n}}}(x)$ and
\begin{equation}
\phi(x)=\left\{\begin{array}{c}
\frac{2 \theta_{max} x}{d}- \frac{\theta_{max}}{2}, 0 \le x < d/2 \\ \\
\pi -\frac{ 2 \theta_{max} x}{d}+ \frac{3\theta_{max}}{2},  d/2 \le x < d.
\end{array} 
\right.
\end{equation}

Clearly, the density at which $N_{SB}$ bifurcates from ordinary nematic (the background phase) has to be compared with that for polar nematic ($N_F$) and A-type smectics: ordinary ($SmA$), ferroelectric ($SmA_F$), and antiferroelectric ($SmA_{AF}$), which are expected for $C_{2v}$-symmetric bow-shaped particles. The negative of the bifurcation density is inversely proportional to the minimum of either the sum or the difference of the cosine coefficients of the Fourier expansion of excluded areas of parallel and antiparallel pairs of the particles divided by 2 (see \cite{karbowniczek}). For the case of smectic analysis the sum and the difference divided by 2 are equal to
\begin{equation}
\begin{split}
(-1+\text{cos}(k-2ak)+\text{cos}(a k \text{cos}(\alpha ))+\text{cos}(k (1-2 a \\
+a\text{cos}(\alpha )))-2 \text{cos}(k (1-2 a+2 a \text{cos}(\alpha )))) \text{tan}(\alpha )/k^2
\end{split}
\end{equation}
and 
\begin{equation}
\begin{split}
-(\text{cos}(k-2 a k)-3 (-1+\text{cos}(a k \text{cos}(\alpha ))+\text{cos}(k(1-2a \\
+a \text{cos}(\alpha ))))+2 \text{cos}(k (1-2 a+2 a \text{cos}(\alpha )))) \text{tan}(\alpha )/k^2,
\end{split}
\end{equation}
\begin{figure}[!htb]
\begin{center}
\includegraphics[width=0.9\columnwidth]{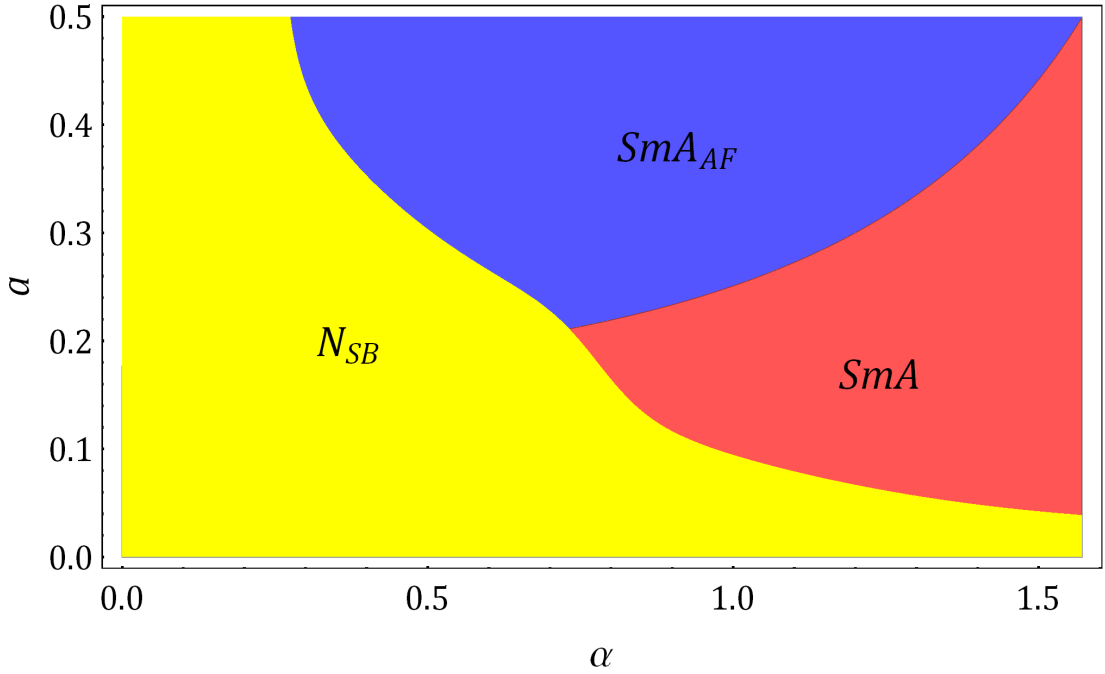}
\end{center}
\caption{Phase diagram of bow-shaped particles for arbitrary $\alpha$ angle and length of the side segment $a$.}
\label{fig1}
\end{figure}
respectively, where $\alpha$ is the angle between the central and the side segment of a particle, $a$ is the length of the side segment, $1-2a$ is the length of the central segment (Fig. \ref{figm1}), and $k$ is the wave vector length. The only two possible smectics bifurcating from the perfect nematic turned out to be the $SmA$ and the $SmA_{AF}$. 
The first structures bifurcating from the perfect nematic are presented in Fig. \ref{fig1} in plane of the $\alpha$ angle and the $a$ length. 
Note, that due to the analytical expressions for the Fourier coefficients the line of the $SmA-SmA_{AF}$ coexistence can be calculated semi-analytically as $a=1/(2+3.664981 \textup{cos}(\alpha))$. For the $N_{SB}$ structure $k=0$ and, since the director field is deformed, $d$ and $\theta_{max}$ are finite. Region of the transition to the $N_{SB}$ is found numerically using optimization with adaptive mesh size.
Densities of the transition to the nematic splay-bend are presented in Fig. \ref{fig2} for several values of $a$. 
\begin{figure}[b]
\begin{center}
\includegraphics[width=0.9\columnwidth]{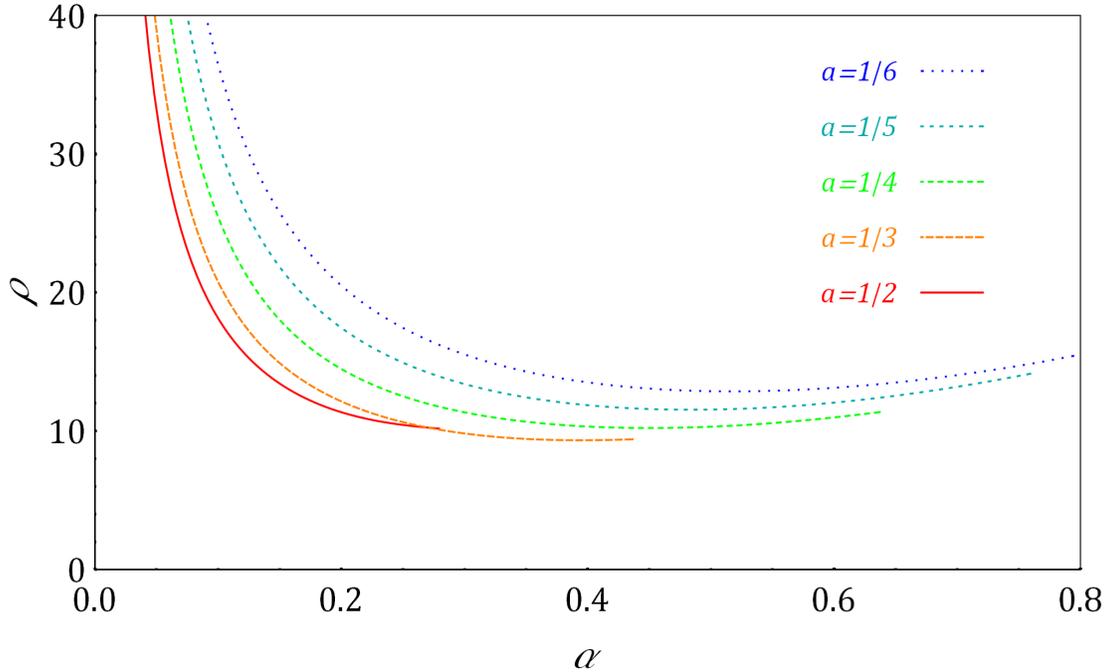}
\end{center}
\caption{Density of phase transition to the $N_{SB}$ for different $\alpha$ angles and lengths of the side segment $a$ (in the legend).}
\label{fig2}
\end{figure}
In Fig. \ref{fig3} periods $d$  of the structure along with $\theta_{max}$ are shown for the same values of $a$ as the lines in Fig. \ref{fig2}. 
\begin{figure}[!htb]
\begin{center}
\includegraphics[width=0.9\columnwidth]{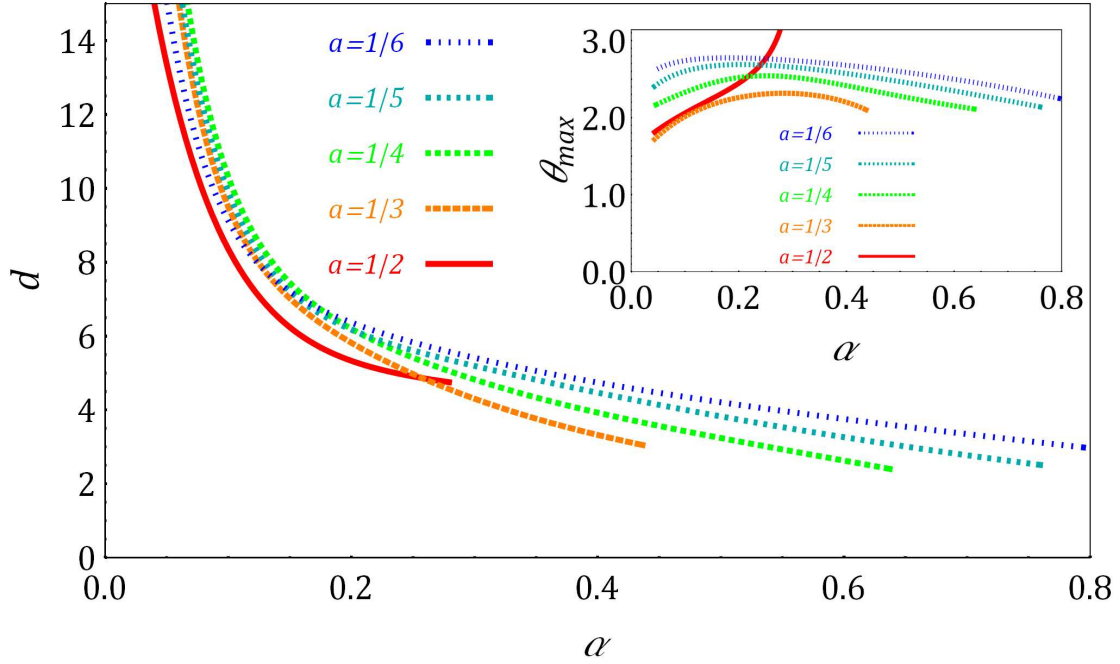}
\end{center}
\caption{Period of the $N_{SB}$ structure as the function of $\alpha$ and $a$ (in the legend). Inset shows $\theta_{max}$ parameter. }
\label{fig3}
\end{figure}
Widths of the lines correspond to the precision of calculations. 
We found that smaller lengths of the side segments of the particles ($a$) imply larger range of the $\alpha$ angles for which the transition to the $N_{SB}$ can occur. However, the downside is that the density of the transition increases. Calculated parameters of the $N_{SB}$ structure are in a qualitative agreement with the results from simulations \cite{tavarone}. 

In conclusion, we confirmed the existence of the nematic splay-bend phase using a simple Onsager-type Density Functional Theory approach and perfect local order approximation. Results are presented here for arbitrary ratios of the length of central and side segments and opening angles of bow-shaped particles. 
\newline

\noindent
This work was supported by Grant No. DEC-2013/11/B/ST3/04247 of the National Science Centre in Poland.


\begin{thebibliography}{9}
\bibitem{tavarone} 
R. Tavarone, P. Charbonneau, and H. Stark, J. Chem. Phys. \textbf{143}, 114505 (2015).
\bibitem{dozov}
I. Dozov, Europhys. Lett. \textbf{56}, 247 (2001).
\bibitem{karbowniczek}
P. Karbowniczek, M. Cie\'sla, L. Longa, and A. Chrzanowska, Liq. Cryst. \textbf{44}, 254 (2017).
\bibitem{meyer} R.B. Meyer, 
\textit{Structural problems in liquid crystal physics. Les Houches Summer School in Theoretical Physics, 1973. Molecular Fluids, eds Balian R, Weil G}, 273, Gordon and Breach, New York (1976).
\end{thebibliography}
\end{document}